# Nonlinear dynamics of a self-mixing thin-slice solid-state laser subjected to Doppler shifted optical feedback

Kenju Otsuka[1] and Seiichi Sudo[2]

[1]*TS³L Research, Yamaguchi 126-7, Tokorozawa, Saitama 359-1145 Japan*
[2]*Department of Physics, Tokyo City University, Tokyo 158-8557 Japan*
*Corresponding author: kenju.otsuka@gmail.com*

Chaotic oscillations of a linearly polarized single longitudinal-mode thin-slice Nd:GdVO$_4$ laser placed in a self-mixing laser Doppler velocity scheme were dynamically characterized in terms of the intensity probability distribution, joint time-frequency analysis, and short-term Fourier transformation of temporal evolutions, and the degree of disorder in the amplitude and phase of the long-term temporal evolutions. The transition from chaotic relaxation oscillations (RO) to chaotic spiking oscillations (SO) was explored via the chaotic itinerancy (CI) regime by increasing the feedback ratio toward the laser from a rotating scattering object. The intensity probability distribution was found to change from an exponential decay in the RO regime to an inverse power law in the SO regime, which manifests itself in self-organized critical behavior, while stochastic subharmonic frequency locking among the two periodicities of RO and SO takes place in the CI regime featuring quantum-noise (spontaneous emission)-induced order in the amplitude and phase of the spiking oscillations. All of the experimental results were reproduced by numerical simulations of a model equation of a single-mode self-mixing solid-state laser subjected to Doppler-shifted optical feedback from a rotating scattering object.

## I. INTRODUCTION

From a fundamental point of view, the laser is a system which allows us to study complex systems far from thermal equilibrium. In particular, it is useful for investigating nonlinear dynamics exhibiting a variety of bifurcations leading to chaos [1].

Chaotic phenomena have been observed in conventional class-B laser systems, where polarization dynamics are adiabatically eliminated. Note that, under most conditions, these systems are stable except within some range of parameter, and each system presents its own deterministic irregular behavior. Period-doubling sequences transitioning to chaotic relaxation oscillations have been reported to occur in solid-state, CO$_2$ and semiconductor lasers with a modulated pump (injection current) or loss at a frequency comparable to the relaxation oscillation frequency [2-4]. Subharmonic bifurcation cascades to chaos and coexisting periodic solutions have observed for various relations between the modulation frequency and frequency of the relaxation oscillation [5].

Subharmonic resonance cascades [6], intermittency [7], period-doubling as well as quasiperiodic routes [8] to chaos have been demonstrated in semiconductor lasers with optical feedback. In fact, a variety of chaotic oscillations have been explored in a semiconductor laser coupled to an external cavity [9], which can be interpreted in terms of delay-driven oscillators involving phase-sensitive interactions between external cavity modes and relaxation oscillations [10].

On the other hand, it is known that laser rate equations can fully describe the motion of a particle in a highly asymmetric laser Toda potential with respect to the ground state, where the damping rate in the rate equations for class-B lasers, $\kappa_{ro}$, increases with the logarithmic photon density, $u(t) = \ln s(t)$, in time, $t$ [11]. In the original Toda oscillator system [12], the damping constant does not depend on $u(t)$. In the absence of a driving force, the particle approaches the ground state by exhibiting damped relaxation oscillations. Chaotic phenomena in the modulated class-B lasers mentioned above can be interpreted in terms of a particle moving within the asymmetric laser Toda potential. The Hamiltonian motion around the ground state (namely, the "soft mode") results from the balance between the damping force, $\kappa_{ro}(du/dt)$, and the driving force from a sinusoidal loss or pump modulation at a frequency comparable to the relaxation frequency. Such periodic "soft-mode" oscillations correspond to sustained periodic relaxation oscillations around the stationary lasing solution, $\bar{s} \propto w - 1$ (w = W/W$_{th}$ : relative pump power normalized by the threshold, W$_{th}$).

By controlling the modulation frequency and amplitude, the periodic "hard-mode" oscillation, which corresponds to regeneration of the first spike in the onset of the relaxation oscillation, was demonstrated to build up from the nonlasing solution of the laser rate equations, $\overline{s_{nl}} \propto 2\varepsilon w/(w - 1)^2 \ll 1$, reflecting quantum (spontaneous emission) noise ($\varepsilon$: spontaneous emission coefficient). Such periodic spike-mode oscillations have been demonstrated in solid-state and semiconductor lasers by using deep sinusoidal loss as well as pump (injection current) modulations [13-16].

If a strong modulation is introduced, chaotic relaxation oscillations (soft mode) and chaotic spiking oscillations (hard mode) will be brought about analogously to a particle moving in the laser Toda potential, where the strong periodic force tends to push the particle irregularly in time because of the unbalanced damping and driving forces. The coexistence of soft-mode and hard-mode attractors is predicted to exist in lasers with delayed incoherent optical feedback, and a physical interpretation based on the laser Toda potential has been given [17].

On the other hand, the effect of noise on nonlinear systems is an intriguing subject from the general viewpoints of nonlinear dynamics and applications. In various devices, an increase in the noise amplitude leads to degradation of the output signal. In nonlinear systems, however, this is not always the case and a finite amount of noise can induce a dynamical state which is more ordered. Examples of such noise-induced order include stochastic excitation of a subharmonic of a periodic modulation signal in a solid-state laser upon modulation of the pump rate by noise and a periodic signal ("stochastic resonance") in the presence of bistability [18] and upon the minimization of pulse interval fluctuations in the intensity of a laser diode with optical feedback when adding noise ("coherence resonance") [19]. However, to date, the studies on noise-induced ordering have been restricted to the effect of "externally applied" artificial noise on the control parameter. In real nonlinear systems, intrinsic quantum noise always exists, and it degrades performance. Here, lasers provide a promising system for investigating the effect of internal intrinsic quantum noise (spontaneous emission) on nonlinear dynamics in a highly asymmetric laser Toda potential.

In this paper, we propose that a thin-slice solid-state laser with coated end mirrors (abbreviated as TS$^3$L) subjected to Doppler-shifted optical feedback from a light scattering object is a well controllable, stable, and affordable system to carry out comprehensive dynamical and statistical studies in a wide modulation range. In the TS$^3$L, the self-mixing modulation effect resulting from interference between the lasing field and feedback field is enhanced because of the large fluorescence-to-photon lifetime ratio, $K = \tau/\tau_p$, on the order of $10^5$-$10^6$, owing to the extremely short photon lifetime [20] and quantum noise affects nonlinear dynamics through the $10^2$-$10^3$ orders of magnitude larger spontaneous emission coefficient as compared with conventional class-B lasers, e.g., solid-state and $CO_2$ lasers, owing to the small active volume. The proposed laser enabled us to perform comprehensive studies on nonlinear dynamics and the effect of quantum noise on modulated class-B lasers in a wide modulation range.

The present work elucidates the nonlinear dynamics hidden in class-B lasers involving soft- and hard-mode oscillations for various relations between the modulation frequency and frequency of the relaxation oscillation and the effect of intrinsic multiplicative quantum noise (spontaneous emission) on hard-mode spiking oscillations that are expected to occur under strong periodic modulation.

By paying special attention to the dynamic characterization of qualitatively different chaotic oscillations, which were observed in the feedback ratio of light scattered toward the laser, the intensity probability distribution was found to change from an exponential decay in the chaotic soft-mode regime to an inverse power law in the chaotic hard-mode regime, via a chaotic itinerancy regime where stochastic frequency locking among two periodicities of soft- and hard-mode oscillations takes place and features a quantum-noise (spontaneous-emission)-induced order in the amplitude and phase. The inverse power law in the intensity probability distributions of the chaotic spiking oscillations, in particular, manifests itself in the laser Toda potential. Over the years, the universality of the inverse power law has attracted particular attention [21,22] for its mathematical properties, and sometimes, its surprising physical consequences in a diverse range of natural and man-made phenomena.

All the experimental results were reproduced in numerical simulations of the model equation of a single mode self-mixing TS$^3$L subjected to Doppler-shifted delayed optical feedback from a moving scattering object.

The paper is organized as follows: the basic properties of TS$^3$L and self-mixing laser Doppler velocimetry are presented in Sec. II. Experimental results on three types of chaotic oscillation are demonstrated for different feedback coefficients in Sec. III. Numerical results are presented in Sec. IV; they confirm that the experimental intensity probability distributions of the chaotic spiking oscillations follow an inverse-power law and that there is quantum-noise-induced order in the chaotic itinerancy regime featuring stochastic frequency locking of soft-mode and hard-mode periodicities. Section V summarizes the results and discusses the physical significance of the observed nonlinear dynamics.

An analogy between class-B laser rate equations and the laser Toda oscillator model is presented in the Appendix as an aid for understanding the soft-mode and hard-mode oscillations described in the main text.

## II. TS$^3$L PROPERTY AND SELF-MIXING LASER DOPPLER VELOCIMETRY SCHEME

Before discussing the characterization of the nonlinear dynamics of the TS$^3$L placed in a state of chaotic oscillations by Doppler-shifted optical feedback scattered from a moving object toward the laser, let us review the basic properties of the Nd:GdVO$_4$ TS$^3$L used in the experiment and self-mixing laser Doppler velocimetry in the weak feedback regime.

### A. Basic properties of thin-slice Nd:GdVO$_4$ laser

The experimental setup is shown in Fig. 1. A nearly collimated lasing beam from a laser diode (LD; wavelength:

808 nm) was passed through an anamorphic prism pair (AP) to transform an elliptical beam into a circular one, and it was focused onto a thin-slice laser crystal by a microscopic objective lens (OL) of numerical aperture NA = 0.5. The laser crystal was a 3 mm-diameter clear-aperture, 300 μm-thick, 3 at%-doped a-cut Nd:GdVO$_4$ whose end surfaces were directly coated with dielectric mirrors M$_1$ (transmission at 808 nm > 95%; reflectance at 1064 nm = 99.8%) and M$_2$ (reflectance at 1064 nm = 99%) and whose Fresnel number was $4 \times 10^4$. The lasing optical spectra were measured by a scanning Fabry-Perot interferometer (SFPI) (Burleigh SA$^{PLUS}$; 2 GHz free spectral range; 6.6 MHz resolution).

The input-output characteristics and an optical spectrum are shown in Figs. 2(a) and 2(b), where linearly polarized single-longitudinal mode oscillations at λ = 1064 nm were observed in the entire pump power region and the intensity fluctuation with respect to the average was measured to be $2\Delta I/I_{av}$ = 5.4%.

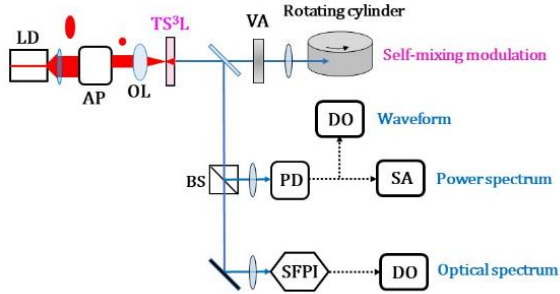

FIG. 1. Experimental apparatus. PD: photo-diode, SFPI: scanning Fabry-Perot interferometer, DO: digital oscilloscope, SA: spectrum analyzer, VA: variable optical attenuator, BS: beam splitter.

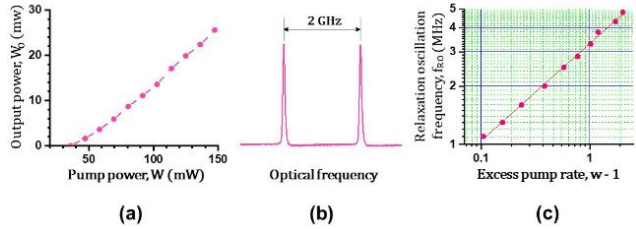

FIG. 2. (a) Input-output characteristics, slope efficiency = 25%. (b) Optical spectrum. Pump power, W = 100 mW. (c) Pump-dependent relaxation oscillation frequency.

The fluorescence-to-photon lifetime ratio, $K = \tau/\tau_p$, which is the key parameter for self-mixing modulations with extreme optical sensitivity, was experimentally evaluated using the pump-dependent relaxation oscillation frequency, $f_{RO} = (1/2\pi)[(w-1)/\tau\tau_p]^{1/2}$, as shown in Fig. 2(c). Assuming $\tau$ = 90 μs for Nd:GdVO$_4$ lasers, the photon lifetime is estimated to be $\tau_p$ = 24 ps. The resultant lifetime ratio is as large as K = 3.75×10$^6$.

## B. Self-mixing laser Doppler velocimetry signals with reduced optical feedback

The self-mixing modulation experiment [23] was carried out using Doppler-shifted light scattered from a rotating cylinder toward the TS$^3$L cavity, as depicted in Fig. 1. Here, the Nd:GdVO$_4$ laser was modulated at $f_D = 2v/\lambda$ because of the self-mixing modulation at the beating frequency between the laser and Doppler-shifted light scattered toward the TS$^3$L cavity (v: moving speed along the lasing axis).

The power spectrum intensity of laser Doppler velocimetry (LDV) signals is proportional to $(\eta K)^2$, where $\eta = |E_b/E_o|$ is the field amplitude feedback ratio ($E_o$: laser output field amplitude, $E_b$: feedback field amplitude toward the laser) [20]. Because of the extremely large K value of $3.75 \times 10^6$, the present TS$^3$L exhibited chaotic oscillations in the absence of a variable optical attenuator, VA, as depicted in Fig. 1. Measurements were carried out using an InGaAs photoreceiver (New Focus 1812, 25 kHz–1 GHz) connected to a digital oscilloscope (Tektronics TDS 3052, DC-500 MHz) and a spectrum analyzer (Tektronix 3026, DC–3 GHz). Typical power spectra for different attenuations are shown in Figs. 3(a)-(c), where each power spectrum was obtained by averaging 100 power spectra measured at intervals of the update, 160 μs. Chaotic oscillations appeared when the round-trip attenuation by the variable attenuator was $T_A \geq -20$ dB. The corresponding power spectrum is shown in Fig. 3(d).

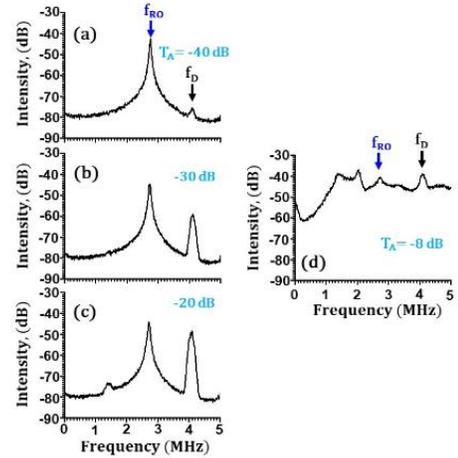

FIG. 3. (a)-(c): Self-mixing LDV signals under weak optical feedback with different roundtrip attenuations, $T_A$. W = 58 mW. (d) Power spectrum corresponding to chaotic oscillations for $T_A = -8$ dB.

## III. DYNAMICAL CHARACTERIZATION OF TS$^3$L PLACED IN CHAOTIC OSCILLATIONS

This section shows the chaotic dynamics observed in different feedback intensity regimes and characterizes the dynamics in terms of intensity probability distributions,

Poincaré sections, return maps of the peak amplitudes and phases, and their degrees of disorder.

### A. Soft-mode chaos and hard-mode chaos

Chaotic relaxation oscillations (namely, "soft-mode" chaos) occurred in the weak optical feedback regime, −20 dB< $T_A$ < −13 dB. A typical chaotic waveform and corresponding fast Fourier transform (power spectrum) are shown in the top panels of Figs. 4(a)-(b), where $f_D$ was tuned to 3 MHz while the relaxation oscillation frequency without modulation was $f_{RO}$ = 2 MHz. The power spectra of such chaotic soft mode chaos have broadened peaks around $f_D$ = 3 MHz and $f_{RO}$ = 2 MHz. On the other hand, chaotic spiking oscillations (namely, "hard-mode" chaos) appeared in the strong optical feedback regime, −9 dB < $T_A$ < 0 dB. An example waveform and corresponding power spectrum are shown in the bottom panels of Figs. 4(a)-(b) below. Note that the spiking frequency was measured from the time series to be on average $f_{SO} \cong$ 1.5 MHz; however, their power spectrum broadened and the $f_{SO}$-peak could not clearly be identified from the power spectrum. This point will be discussed in section **III**-**B**.

The intensity probability distributions for soft- and hard-mode chaos are shown in Fig. 4(c), where fitting lines and Pearson product-moment correlation coefficients, R, are depicted. From repeated acquisition of chaotic time series for various values of $f_D$ and $f_{RO}$ in the regime of $f_D \geq f_{RO}$, the intensity probability distribution was concluded to obey an exponential decay $P(s) \propto e^{-\alpha s}$ for soft-mode chaos and an inverse power law, $P(s) \propto s^{-\beta}$ for hard-mode chaos. The scaling parameter typically lay in the range 0.9 < α, β < 2, while |R| > 0.99.

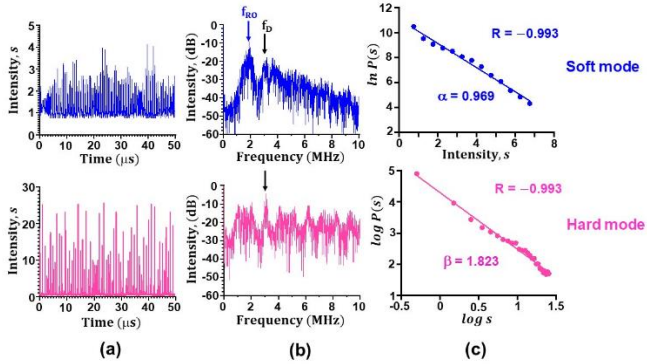

FIG. 4. (a) Temporal evolutions of soft-mode and hard-mode chaos. The output voltage was calibrated by using DC bias output. (b) Corresponding power spectra. (c) Intensity probability distributions. The base of the logarithmic functions is e (Euler's Number) and 10 for soft mode and hard mode, respectively. W = 50 mW.

The universality of the inverse power law for chaotic spiking oscillations is surprising because the intensity probability distributions of spike-pulse waveforms in a periodic spiking oscillation deviate from an inverse power law, as will be addressed in the Appendix I. In short, spike-pulses with random peak intensities and pulse widths are self-organized such that an inverse power law $P(s) \propto s^{-\beta}$ is established for overall spike-pulses in the long-term evolutions. This suggests that the laser Toda potential system driven by a strong force exhibits self-organized critical behaviour, while it has been argued that earthquakes, landslides, forest fires, sizes of power outages and species extinctions are examples of self-organized criticality in nature [24]. This point will be discussed in the Appendix II.

Figures 5(a) and 5(b) show the Poincaré sections $[s, \dot{s}]$ and return maps of the peak intensities and the time interval between peaks, $[s_{p,i}, s_{p,i+1}]$ and $[t_{p,i}, t_{p,i+1}]$, for soft- and hard-mode chaotic oscillations. The Poincaré sections of the soft-mode and hard-mode chaos show qualitatively different topologies and coexist in the phase space. The standard deviations of the peak intensities, $R_A$ = A/<A>, and of the time interval between peaks, $R_T$ = T/<T>, are a measure of disorder in the amplitude and phase. From repeated experiments, $R_A$ and $R_T$ were evaluated be 0.31 and 0.25 on average for the soft-mode chaos, while these values were found to increase to $R_A$ = 0.39, $R_T$ = 0.31 on average for the hard-mode chaos. This suggests a larger degree of disorder in spiking chaos.

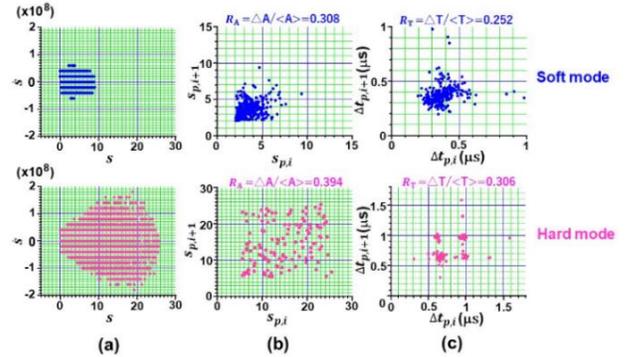

FIG. 5. (a) Poincaré sections of soft- and hard-mode chaos. (b) Return maps of peak intensities. (c) Return maps of time interval between peaks.

### B. Chaotic itinerancy: alternating appearance of soft-mode and hard-mode chaos

When the optical feedback was increased beyond the range of soft-mode chaos, i.e., $T_A \geq -13$ dB, bursts of spiking chaos appeared and the survival times of the chaotic relaxation oscillations, $t_{RO}$ = (sum of dwell times in relaxation oscillations)/(total time series) decreased with increasing optical feedback, leading to the chaotic spiking oscillations shown in Fig. 4 above $T_A \leq -9$ dB. In short, self-induced switching emerged in which there were chaotic relaxation oscillations and chaotic spiking oscillations at the boundary between soft-mode and hard-mode chaos. This unusual behavior has been observed in $LiNdP_4O_{12}$ $TS^3L$ modulated at harmonic frequencies of $f_{RO}$ [25], where numerical simulations proved that self-induced switching between the ruins of soft and hard-mode attractors takes place deterministically even in the absence of quantum (spontaneous emission) noise. Therefore, switching behavior manifests itself in the so-called "chaotic itinerancy" (namely, CI) frequently observed in vast complex systems [26-32].

The concept of CI refers to dynamical behavior in which the system itinerates over the "ruins" of localized chaotic attractors in some irregular way, while chaotic dynamics enables the system to form *easy switching paths* among the localized chaotic attractors [26]. During this itinerancy, the orbits visit a neighborhood of an attractor ruin with a relatively regular and stable motion, for relatively long times, and then the trajectory jumps to another attractor ruin of the system. In addition, in the present TS³L system subjected to self-mixing modulations, the laser's quantum noise has an intriguing effect on the itinerant behavior.

Figure 6(a) shows chaotic waveforms for different round-trip optical attenuations, $T_A$, observed at the pump power, W = 50 mW. As shown in Fig. 6(b), the survival time $t_{RO}$ decreased according to an inverse power law, i.e., $t_{RO} \propto T_A^{-\gamma}$. The distinct power spectrum and intensity probability distribution observed for $T_A = -11.5$ dB are shown in Fig. 6(c). It is interesting that a clear peak appeared at the spiking frequency, $f_{SO} = f_D/2 = 1.5$ MHz, while it was broadened for the chaotic hard-mode spiking waveforms (Fig. 4). Furthermore, the relaxation oscillation frequency in the soft-mode chaos [Fig. 4(a)], which is depicted by the dashed line, shifted from 2 MHz to $f_{RO} = (3/4) f_D = 2.25$ MHz toward $f_D$. Such a nonlinear frequency pulling of the relaxation oscillation frequency without modulations toward the modulation frequency $f_D$ has often been identified in the CI regime. As for the intensity probability distribution, the peculiar "slope" appeared to correspond to a quiet region of $s(t)$ between the upper bound of soft-mode chaos intensity fluctuations and the lower bound of hard-mode chaotic fluctuations.

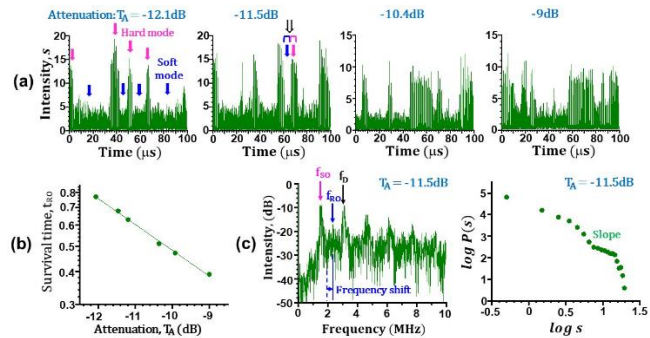

FIG. 6. Chaotic itinerancy observed at W = 50 mW. (a) Temporal evolutions for different attenuations, $T_A$. (b) Survival time of soft-mode chaos versus attenuation. (c) Power spectrum and intensity probability distribution for $T_A = -11.5$ dB.

The power spectrum shown in Fig. 6(c) suggests nonlinear frequency locking among two periodicities $f_{RO}$ and $f_{SO}$ in the form of $[f_{SO} : f_{RO}] = [2 : 3]$ through the self-mixing modulation at $f_D$ in the CI regime. To clarify this situation, let us examine the dynamics occurring in the switching regions indicated by ⇓ in Fig. 6 for $T_A = -11.5$ dB through a joint time-frequency analysis (JTFA) of time series [33]. The results are shown in Fig. 7(a), while the short-term Fourier transform power (power spectra) around RO and SO chaos indicated by ↓ are shown in Fig. 7(b). It is obvious that a [2 : 3] type of frequency locking is established in the switching region between the soft- and hard-mode time series during the temporal evolutions, as indicated by ⇑ at $3f_{SO} = 2f_{RO}$ in Fig. 7(b), On the other hand, the degree of disorder in the amplitude and phase in the overall spiking time series is greatly lowered to $R_A = 0.193$ and $R_T = 0.062$, as shown in Fig. 7(c). This implies that spiking chaos oscillations are tamed by frequency locking among the two periodicities, $f_{SO}$ and $f_{RO}$. The appearance of such tamed spiking oscillations is presumably considered to involve quantum noise, i.e., spontaneous emission, since spiking oscillations are born from a nonlasing solution which depends on the spontaneous emission coefficient, as mentioned in the Introduction. The quantum-noise-induced order will be addressed later in numerical simulations.

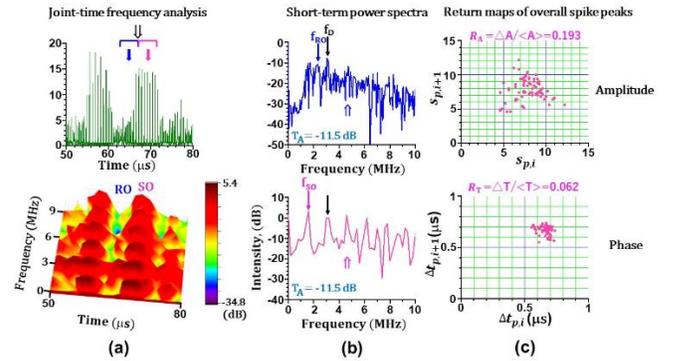

FIG. 7. (a) Zoomed-in views of time series for $T_A = -11.5$ dB and its joint time-frequency analysis. (b) Short-term Fourier transform (power spectra) around switching point. (c) Return maps for hard-mode spiking peaks.

Such a [p : q] frequency locking accompanied by formation of an *easy switching path* among the ruins of soft- and hard-mode attractors [26] was identified in the CI regime for various sets of $f_D$ and $f_{RO}$. Experimental results exhibiting [3 : 4] and [1 : 2] frequency locking, which feature tamed spiking oscillations, are shown in Figs. 8 and 9, respectively.

As for Fig. 8, spiking oscillations were excited at $f_{SO} = f_D/2$ and exhibited subharmonic locking with $(3/4) f_{RO}$, as indicated by ⇑, while $f_{RO}$ in the soft-mode chaos was found to shift toward $f_D$ in the CI regime, similarly to Fig. 7.

As for Fig. 9, $f_D$ was tuned close to $f_{RO}$, while $f_{RO}$ was locked to $f_D$ and exhibited subharmonic locking with $f_{SO} = (1/2)f_D = (1/2)f_{RO}$. Tamed spiking oscillations with decreased disorder in amplitude and phase, $R_A$ and $R_T$, were brought about in both cases, as depicted in Figs. 8(c) and 9(c).

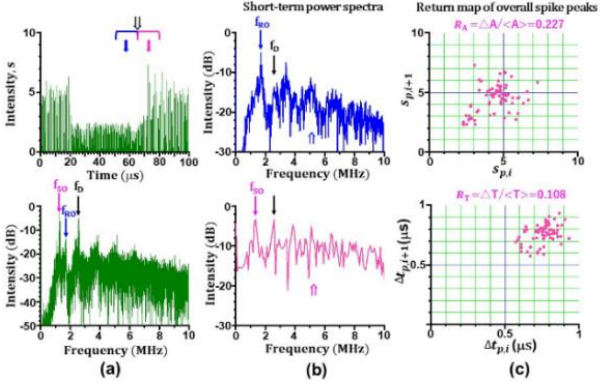

FIG. 8. Chaotic itinerancy featuring [3 : 4] frequency locking. W = 46 mW. $T_A$ = −10 dB.

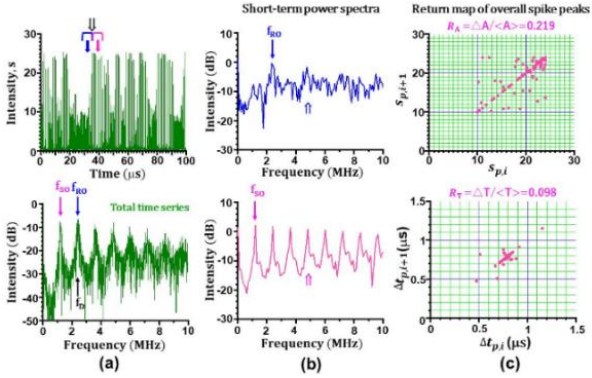

FIG. 9. Chaotic itinerancy featuring [1 : 2] frequency locking. W = 70 mW. $T_A$ = −12 dB.

## IV. NUMERICAL RESULTS

The numerical results presented below support the experimental observations and conjectures discussed so far.

The generalized dynamical equations for self-mixing lasers can be obtained by extending the Lang-Kobayashi equations [9] to include Doppler-shifted feedback of light scattered from a moving target [20]:

$$dN(t)/dt = \{w - 1 - N(t) - [1 + 2N(t)]E(t)^2\}/(K/2), \quad (1)$$

$$dE(t)/dt = N(t)E(t) + \eta E(t - t_D)\cos\Psi(t) + \{2\varepsilon[N(t) + 1]\}^{1/2}\xi(t), \quad (2)$$

$$d\phi(t)/dt = \eta[E(t - t_D)/E(t)]\sin\Psi(t), \quad (3)$$

$$\Psi(t) = \Omega_D t - \phi(t) + \phi(t - t_D). \quad (4)$$

Here, $E(t) = (g\tau)^{1/2}\mathbf{E}(t)$ is the normalized field amplitude, and $N(t) = g\mathbf{N}_{th}\tau_p(\mathbf{N}(t)/\mathbf{N}_{th} - 1)$ is the normalized excess population inversion, where $\mathbf{N}_{th}$ is the threshold population inversion. $g$ is the differential gain coefficient, where gain is defined as $\mathbf{G} = \mathbf{G}_{th} + g(\mathbf{N}(t) - \mathbf{N}_{th})$. $w = \mathbf{W}/\mathbf{W}_{th}$ is the relative pump rate normalized by the threshold, $\phi(t)$ is the phase of the lasing field, $\Psi(t)$ is the phase difference between the lasing and the feedback field, and $\eta$ is the amplitude feedback ratio. $\Omega_D = \omega_D/\kappa$ is the normalized instantaneous frequency shift of the feedback light from the lasing frequency. $t$ and $t_D$ are the time and delay time normalized by the damping rate of the optical cavity $\kappa = 1/(2\tau_p)$. The last term of Eq. (2) includes quantum (spontaneous emission) noise, where $\varepsilon$ is the spontaneous emission coefficient and $\xi(t)$ is Gaussian white noise with zero mean, and the value $<\xi(t)\,\xi(t')> = \delta(t - t')$ is δ-correlated in time. In the short-delay limit, Equations (1) – (4) reduce to the laser rate equations, which are given in the Appendix I.

### A. Soft-mode chaos and hard-mode chaos

The temporal evolutions were calculated by using parameters corresponding to the experimental results of the soft-mode and hard-mode chaos shown in Figs. 4(a) and 4(b), assuming $w = 1.38$, $K = 3.75 \times 10^6$, $\varepsilon = 10^{-8}$, $\Omega_D = 9 \times 10^{-4}$ and $t_D = 10$. Numerical results representing soft-mode and hard-mode chaos are shown in Figs. 10(a) and 10(b), assuming $\eta = 1.65 \times 10^{-4}$ and $8 \times 10^{-4}$, respectively. Here, the time scale was returned to real time by using $\tau_p = 24$ ps. The threshold intensity feedback ratio for chaotic oscillations was found to decrease as $\varepsilon$ increases. Numerical waveforms and the corresponding power spectra match the experimental results in Figs. 4(a) and 4(b) remarkably well. In particular, the peculiar power spectrum for chaotic spiking reproduces the experimental one in detail, as shown in the bottom panel of Fig. 4(b). The calculated intensity probability distributions for chaotic relaxation and spiking oscillations are shown in Fig. 10(c), where correlation coefficients are evaluated to be R = −0.996 and −0.992, respectively. These values are also close to the experimentally obtained R = −0.993 for both chaotic relaxation and spiking oscillations in Fig. 4(c).

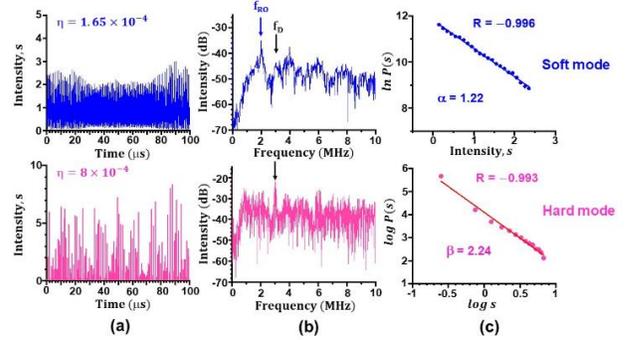

FIG. 10. Numerical results on soft- and hard-mode chaos. $w = 1.38$, $K = 3.75 \times 10^6$, $\varepsilon = 10^{-8}$, $\Omega_D = 9 \times 10^{-4}$ and $t_D = 10$.

The calculated Poincaré sections and return maps of the peak intensities and their time intervals, $[s_{p,i}, s_{p,i+1}]$ and $[t_{p,i}, t_{p,i+1}]$, for chaotic relaxation and spiking oscillations are shown in Figs. 11(a) and 11(b). The associated standard deviations of the peak intensities, $R_A = A/<A>$, and the time interval between peaks, $R_T = T/<T>$, were $R_A = 0.31$, $R_T = 0.21$ for chaotic relaxation and $R_A = 0.43$, $R_T = 0.34$ for chaotic spiking oscillations. These degrees of disorder are close to the average values of $R_A = 0.31$, $R_T = 0.25$ for chaotic relaxation oscillations and $R_A = 0.39$, $R_T = 0.31$ for chaotic spiking oscillations obtained in the experiment (Fig. 5).

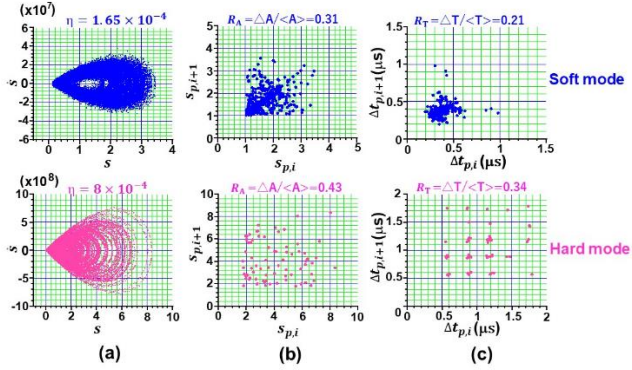

FIG. 11. Calculated Poincaré sections and return maps of peak intensities and time intervals between peaks corresponding to Fig. 10.

### B. Chaotic itinerary: alternating appearance of soft-mode and hard-mode chaos

An example of the calculated temporal evolutions exhibiting alternating chaotic relaxation and spiking oscillations in the CI regime is shown in Fig. 12(a), together with the intensity probability distribution that features a "slope", assuming $\eta = 1.8 \times 10^{-4}$. The other parameters are the same as those for the chaotic soft- and hard-mode oscillations shown in Fig. 10.

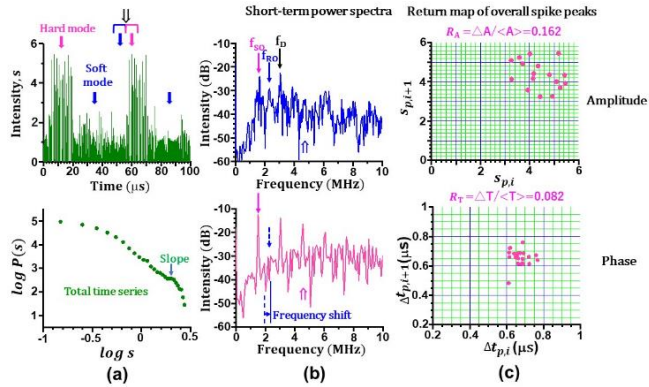

FIG. 12. Numerical results of chaotic itinerancy. $\eta = 1.8 \times 10^{-4}$. Other parameters are the same as those for Fig. 10.

The short-term power spectra of soft- and hard-mode chaos in the switching region indicated by ⇓ are shown in Fig. 12(b). The clear peak corresponding to spiking oscillations at $f_{SO} = f_D/2$ as well as a nonlinear shift in the relaxation oscillation frequency from 2 MHz to $f_{RO} = (3/4)f_D$ = 2.25 MHz toward $f_D$ match the experimental results shown in Fig. 7.

The calculated return maps of the peak intensities and their time intervals, $[s_{p,i}, s_{p,i+1}]$ and $[t_{p,i}, t_{p,i+1}]$, for the overall hard-mode spiking time series are shown in Fig. 12(c). The associated standard deviations of the peak intensities, $R_A = A/<A>$, and the time interval between peaks, $R_T = T/<T>$ were $R_A = 0.162$ and $R_T = 0.082$ for tamed chaotic spiking oscillations. These values are close to the average experimental values of $R_A = 0.193$ and $R_T = 0.062$ in the experimental results in the CI regime.

The parametric excitation of spiking oscillations due to harmonic frequency modulations at $f_D = 2 \times f_{SP}$ as well as resonant relaxation oscillations featuring a nonlinear $f_{RO}$ pulling toward $f_D$, which reproduce the experimental results shown in **III-B**, are considered to encourage subharmonic frequency locking of the two periodicities of the soft mode and hard mode in the CI regime. In addition, periodic spike-mode oscillations due to harmonic frequency injection-current and pump modulations were demonstrated in semiconductor and solid-state lasers, respectively [15, 16].

### C. Numerical analysis of the effect of spontaneous emission noise on chaotic itinerancy

Let us examine the effect of spontaneous emission noise on self-induced switching among chaotic relaxation and spiking oscillations on the basis of the numerical simulations.

Self-induced switching among the ruins of soft and hard-mode attractors was already proved to take place deterministically even in the absence of the quantum (spontaneous emission) noise term in Eq. (2) [25]. On the other hand, the experimental results and numerical reproductions shown in the previous sections strongly suggest that quantum (spontaneous emission) noise affects the deterministic chaotic itinerancy phenomenon, i.e., taming of spiking oscillations presumably associated with stochastic frequency locking among two periodicities of soft and hard modes, i.e., $f_{SO}$ and $f_{RO}$.

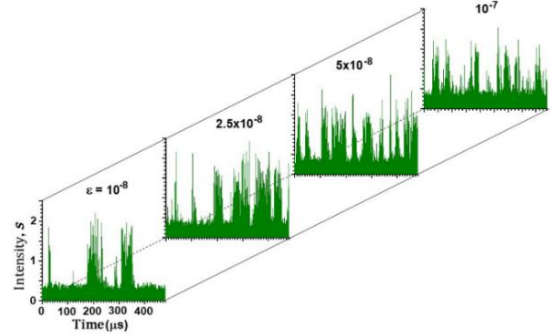

FIG. 13. Effect of spontaneous emission noise on chaotic itinerancy. $w = 1.1$, $K = 3.75 \times 10^6$, $\eta = 2.5 \times 10^{-4}$ and $\Omega_D = 2 \times 10^{-4}$.

Here, numerical simulations were carried out for various values of the spontaneous emission coefficient, ε, using Eqs. (1)-(4). Example results are shown in Fig. 13, assuming w = 1.1, $K = 3.75 \times 10^6$, $\eta = 2.5 \times 10^{-4}$ and $\Omega_D = 2 \times 10^{-4}$, where the total integration time is $10^7$. The spontaneous emission coefficient is given by $\varepsilon = c\sigma\tau/\pi w_o nL$, where c is the velocity of light, σ is the stimulated emission cross section, $w_o$ is the lasing beam spot size averaged over the cavity length L, and n is the refractive index. Large spontaneous emission coefficients, on the order of $10^{-8} \sim 10^{-7}$, are estimated for the short-cavity TS³Ls with L ≤ 1 mm. Figure 13 indicates that the survival time of soft-mode chaotic relaxation oscillations, $t_{RO}$ = (sum of dwell times in soft-mode oscillations)/(total integration time), depends on the strength of the spontaneous emission noise, i.e., ε.

To clarify the effect of ε on $t_{RO}$, repeated numerical experiments were performed for different values of ε. The survival time, $t_{RO}$, averaged over ten data sets of $10^7$ points each is plotted as a function of ε in Fig. 14(a), where the critical intensity that distinguishes hard-mode (spiking) chaos from soft-mode (relaxation oscillation) chaos was set to $I_c = 0.5$ for all datasets.

The survival time, $t_{RO}$, reaches a minimum value and increases afterwards as ε increases, showing increased intensity fluctuations. In addition, the degree of disorder was evaluated in terms of the standard deviations of the peak intensities, $R_A$ = A/<A>, and the time interval between peaks, $R_T$ = T/<T> for the overall time series of the hard-mode spiking oscillations. It is obvious that the degree of disorder reaches a minimum value at a certain ε, corresponding to the minimum for $t_{RO}$ around $\varepsilon = 5 \times 10^{-8}$, as shown in Fig. 14(a).

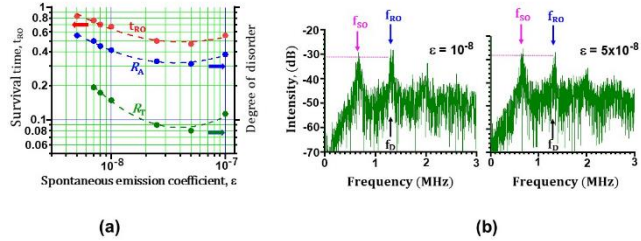

FIG. 14. (a) Survival time of soft-mode chaos and degree of amplitude disorder versus spontaneous emission coefficient corresponding to Fig. 13. (b) Power spectra near the optimum ε of $5 \times 10^{-8}$.

These results imply that an optimum value of ε exists at which chaotic spiking oscillations are mostly tamed over time through stochastic frequency locking among the periodicities of $f_{RO}$ (relaxation oscillation) and $f_{SO}$ (spiking oscillation). The power spectra for $\varepsilon = 1 \times 10^{-8}$ and $5 \times 10^{-8}$ are shown in Fig. 14(b). They indicate that the relaxation oscillation frequency in the free-running condition ($f_{RO}$ = 1.1 MHz) is pulled toward the self-mixing modulation frequency and frequency locked to $f_{RO} = f_D$ = 1.32 MHz. Then, chaotic spiking oscillations are excited at $f_{SO}$ = (1/2)$f_{RO}$. The $f_{SO}$-peak intensity was enhanced mostly at the optimum value of ε, around $5 \times 10^{-8}$ in this case. As ε increases beyond the optimum value, the system tends to develop chaotic spiking oscillations with increased disorder.

The theoretical results shown in Figs. 12~14 verify the effect of quantum (spontaneous emission) noise on deterministic chaotic itinerary phenomena and explain the quantum-noise-induced order through stochastic frequency locking of two periodicities of $f_{RO}$ and $f_{SO}$ observed in the experiments. Moreover, such noise-induced ordering did not arise by adding Gaussian white noise to the pump rate in the form of w + δ ξ(t) in the simulation, where δ is the noise strength.

## V. CONCLUSIONS

In summary, systematic investigations were performed on the nonlinear dynamics of a laser-diode-pumped thin-slice Nd:GdVO₄ laser (TS³L) subjected to self-mixing modulation with a Doppler-shifted field from a moving scattering object. The highly sensitive self-mixing modulation effect inherent to a thin-slice solid-state laser with an extremely large fluorescence-to-photon lifetime ratio as well as its large spontaneous emission coefficient led to the observation of the generic dynamics hidden in modulated class-B lasers. These dynamics include self-organized critical behavior of chaotic spiking oscillations, which exhibit intensity probability distributions obeying an inverse power law, chaotic itinerancy (CI) among the ruins of relaxation oscillations and spiking oscillations around frequencies of $f_{RO}$ and $f_{SO}$, and taming of chaotic spiking oscillations, with an enhanced degree of order in amplitude and phase, through quantum (spontaneous emission) noise mediated stochastic subharmonic frequency locking among two periodicities, $f_{RO}$ and $f_{SO}$ (<$f_{RO}$), in the CI regime.

All the experimental results were correctly reproduced in numerical simulations of a model equation of self-mixing class-B lasers including spontaneous emission noise. The theoretical approach toward the universality of the inverse power law inherent to chaotic spiking oscillations in laser Toda potential remains as an interesting future task.

## APPENDIX I: LASER TODA OSCILLATOR MODEL

This Appendix reviews the analogy between laser rate equations and the Toda oscillator system as an aid for understanding the soft-mode and hard-mode oscillations described in the main text.

The laser rate equations for class-B lasers subjected to periodic loss modulation are given by:

$$dn/dt = w - n + ns, \qquad (5)$$

$$ds/dt = K[\{n - (1 + m\cos\omega_D t)\}s + \epsilon n]. \qquad (6)$$

Here, *w* is the pump power normalized by the threshold pump power, *n* is the population inversion density normalized by the threshold, *s* is the normalized photon density, m = 2η is the modulation amplitude, $\omega_D = 2\pi\tau f_D$ is the normalized angular modulation frequency, ε is the spontaneous emission coefficient, and time is normalized by τ. In the short delay limit, i.e., $t_D \to 0$, Eqs. (1)-(4) in section **IV** reduce to the laser rate equations (5)-(6) with a loss modulation term. These rate equations are considered to describe the fundamental dynamics of self-mixing lasers [20].

To provide physical insight into relaxation oscillations (soft mode) and spiking oscillations (hard mode) described in the main text, let us introduce the Toda potential for laser rate equations (5) and (6), which include a spontaneous emission term. The dynamics of the photon density can be understood in analogy with the motion of a particle in the following laser Toda potential, V, by making a logarithmic transformation, $u(t) \equiv \ln s(t)$:

$$d^2u/dt^2 + \kappa_{ro}(du/dt) + \partial V/\partial u = F_D, \quad (7)$$

$$V = K[w\varepsilon e^{-u} + (e^u - 1)(1 + m\cos\omega_D t) - (w-1)u - w\varepsilon], \quad (8)$$

$$\kappa_{ro} = \varepsilon[K(1 + m\cos\omega_D t) + \frac{du}{dt}]/(e^u + \varepsilon) + 1 + e^u, \quad (9)$$

$$F_D = Km\omega_D \sin(\omega_D t) - Km\cos(\omega_D t). \quad (10)$$

Equations (8)-(9) contain an additional $\cos(\omega_D t)$ component, which is responsible for parametric excitation of the nonlinear laser oscillator subjected to loss modulation.

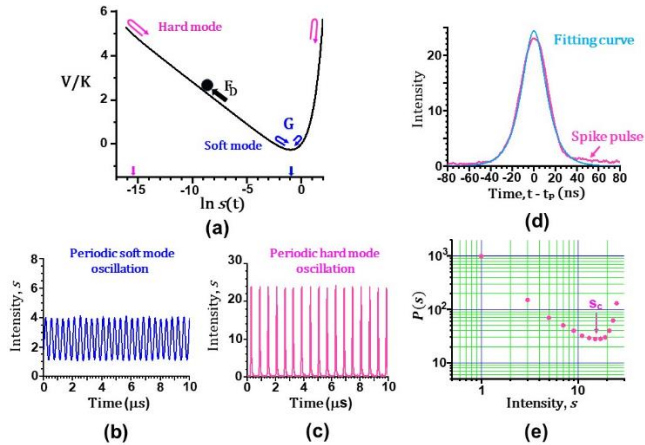

FIG. 15. (a) Particle motion corresponding to soft- and hard-mode oscillations in laser Toda potential. (b) Periodic soft-mode oscillation. (c) Periodic hard-mode oscillation. (d) Magnified view of single spike-pulse and hyperbolic-function fitting curve. (e) Intensity probability distribution of periodic spiking oscillations shown in (c). W = 57 mW.

The calculated laser Toda potential is shown in Fig. 15(a), assuming **m = 0**, w = 1.38 and ε = **1** ×10⁻⁸. The particle moves within a highly asymmetric laser Toda potential with a complicated damping rate, $\kappa_{ro}$. Here, the damping rate increases as the photon density, $s(t) = e^{u(t)}$, increases, whereas it does not depend on $u(t)$ in the original Toda oscillator [12].

Without a driving force, $F_D$, in Fig. 15(a), the particle approaches the ground state, G, which corresponds to the lasing stationary solution of Eqs. (5)-(6) without modulation, $\overline{s_l}$ = w – 1, and shows damped relaxation oscillations.

The Hamiltonian motion around the ground state, i.e., periodic relaxation oscillations (soft mode), is established if a periodic driving force is applied at the relaxation oscillation frequency, $f_{RO}$, such that the damping force, $\kappa_{ro}(du/dt)$, balances the periodic driving force, $F_D$. In addition to the soft mode, a spike-like waveform that builds up from the nonlasing stationary solution of Eqs. (5)-(6), $\overline{s_{nl}}$ = $2\varepsilon w/(w-1)^2$ <<1, within the asymmetric potential is also expected to manifest itself in the large signal regime by tuning the strength and frequency of the driving force. In fact, periodic spiking oscillations (hard-mode) were realized in semiconductor lasers through the use of deep injection current modulation [16] and in solid-state lasers through the use of deep loss or pump modulation at $f_{SP}$ (< $f_{RO}$) [13-15].

In the present self-mixing modulation scheme, periodic relaxation and spiking oscillations were obtained by tuning the modulation frequency, $f_D$, and feedback rate, $T_A$, as shown in Fig. 15(b) and 15(c), respectively. The modulation frequency $f_D$ was tuned close to $f_{RO}$ = 2.7 MHz for the periodic soft-mode oscillation, while the periodic hard-mode oscillation was obtained when $f_D$ was set to $f_{SP}$ = 1.7 MHz. The optical feedback rate was adjusted appropriately to preserve sustained periodic oscillations in both cases. The pulse width (FWHM) of the periodic spikes was measured to be 30 ns.

Note that the spike-pulse waveform is well fitted by the following hyperbolic function:

$$s(t) = s_p \text{sech}^2\left(\sqrt{\frac{s_p}{2\tau\tau_p}}(t - t_0)\right), \quad (11)$$

similarly to Carson's giant pulse model [34]. $t_o$ is the time at which the peak photon number occurs. A magnified view of a single spike-pulse waveform in Fig. 15(c) and the hyperbolic fitting curve are shown in Fig. 15(d), assuming τ = 90 μs and $\tau_p$ = 24 ps. A large coefficient of determination of $R^2$ = 0.99 is attained in this case. The asymmetric nature in the pulse shape in Fig. 15(d) is considered to arise from larger initial population inversions in the onset of a spike pulse, with a faster rise time and a slower decay time [35]. Furthermore, it is interesting that the spike-pulse waveform is formally equivalent to the following 1-soliton solution of the Toda oscillator system [36]:

$$\Phi_n(t) = (sinh^2 k)sech^2[(sinh k)(t-t_0)+kn], \quad (12)$$

which is found by normalizing the time by $\sqrt{2\tau\tau_p}$ and setting $s_p$ in Eq. (11) to $sinh^2 k$, where $k$ is the spring constant and $n$ is an integer.

The intensity probability distribution $P(s)$ for the periodic spiking oscillation corresponding to Fig. 15(c) is shown in Fig. 15(e). $P(s)$ does not obey an inverse power law as expected from Eq. (11), while it reaches a minimum at the inflection point, $s_c$, which is expressed by the following equation:

$$s_c = s_p sech^2\left(arctanh\frac{1}{\sqrt{3}}\right) = \frac{2}{3}s_p. \quad (13)$$

$P(s)$ increases monotonically in the region $s > s_c$.

In the chaotic spiking oscillations shown in III-A, chaotic spike-pulses with different peak intensities, $s_{p,i}$, and pulse widths (FWHM), $\Delta\tau_{p,i}$, were fitted by the hyperbolic function given by Eq. (11) similarly to Fig. 15(d), while $\Delta\tau_{p,i}$ increases with decreasing $s_{p,i}$. This suggests that the self-organized critical behavior in chaotic spiking oscillations obeys an inverse power law, where chaotic spike-pulses are organized in the laser Toda oscillator system such that the overall intensity probability distribution for chaotic spike-pulses during long-term evolutions obeys an inverse power law.

Finally, let us show an intriguing functional transition of spike-pulse waveform from the hyperbolic function given by Eq. (11) for hard-mode spiking chaos to the Gaussian function for quantum-noise-mediated tamed spiking chaos in the CI regime, which has not been identified in the original Toda oscillator system without noise.

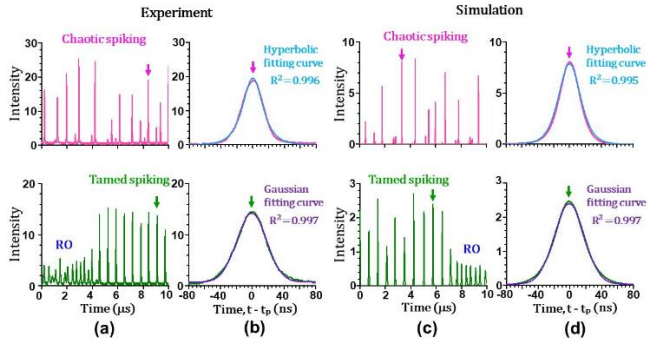

FIG. 16. Functional transition of spike-pulse waveforms from hyperbolic to Gaussian associated with quantum-noise-induced frequency locking in the chaotic itinerancy. (a), (b): Experimental results. (c), (d): Numerical results.

Zoomed-in views of the experimental chaotic spiking oscillation and tamed spiking oscillation in the CI regime corresponding to Fig. 4(a) and Fig. 7(a) are shown in Fig. 16(a). The corresponding magnified waveforms of spike-pulses indicated by the arrows are shown in Figs. 16(b), together with hyperbolic and Gaussian fitting curves. Figures 16(c)-(d) show zoomed-in views of numerical results corresponding to Fig. 10(a) and 12(a). The numerical results reproduce the experimental ones. The non-trivial functional transition from hyperbolic to Gaussian is apparent, featuring large coefficients of determination, $R^2$. All spike-pulses with different peak intensities and tamed spiking oscillations were fitted by hyperbolic and Gaussian functions, respectively.

Although a concrete theoretical explanation has been left as a future task, we showed analytical expressions for a particle moving in a laser Toda potential subjected to self-mixing modulation. The parametric excitation of the nonlinear oscillator responsible for subharmonic frequency locking of two periodicities through quantum noise as well as the mathematical equivalence between the spike-pulse (hard mode) and soliton solutions of the Toda lattice have been explored. Moreover, the non-trivial functional transition of spike-pulse waveforms from hyperbolic to Gaussian in the chaotic itinerancy was reproduced theoretically. These results suggest the robustness of the model equations of a TS$^3$L subjected to self-mixing modulation, Eqs. (1)-(4), and the corresponding laser Toda oscillator expressed by Eqs. (7)-(10).

## APPENDIX II: STATISTICAL GRAPHICS OF CHAOTIC SPIKE-PULSES

Finally, we show the intriguing relation between three quantities of peak intensity, $s_p$, pulse width, $\Delta\tau_p$, and pulse energy (area), $E_p$, of spikes, which is established behind the inverse-power law of intensity probability distributions. Each spike-pulse waveform is approximated by a hyperbolic function, while it exhibits an asymmetric nature with a long-time tail depending on the peak intensity. We calculated these three quantities for 165 spike-pulses of chaotic spiking oscillations shown in Fig. 4(a) by using ORIGINPRO software, as depicted in the top panel of Fig. 17(a). The 3D statistical graphics of spike-pulses and its projections are shown in the bottom panel of Fig. 17(a). Individual pulse energies are self-organized to lie on the parabolic surface given by $E_p = E_{p,0} + as_p + b\Delta\tau_p + cs_p^2 + d\Delta\tau_p^2$ with a large coefficient of determination, $R^2 = 0.996$, as shown in Fig. 17(b).

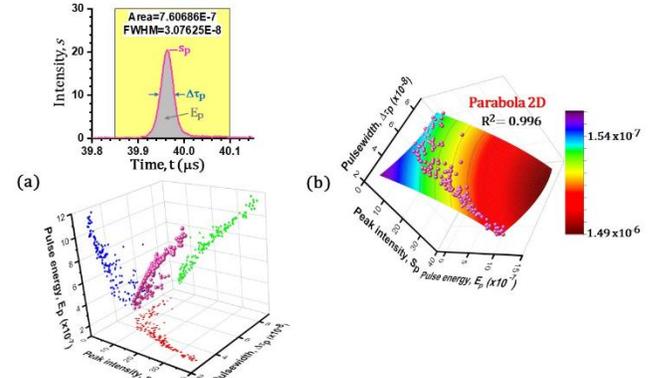

FIG. 17. Statistical graphics of chaotic spike-pulses. Fitting parameters: $E_{p,0}= -1.184 \times 10^{-6}$, a $= 5.782\times 10^{-8}$, b $=36.281$, c $=-5.274\times 10^{-10}$, d $= -2.623\times 10^{8}$.

[1] F. T. Arecchi and R. G. Harrison (eds.): Instabilities and Chaos in Quantum Optics, (Springer-Verlag, 1987).
[2] W. Klische, H. R. Telle, and C. O. Weiss, Opt. Lett. **9**, 561 (1984).
[3] P. Glorieux, Journal de Physique Colloques **48**, C7-433 (1987).
[4] L. Chusseau, E. Hemery, and J-M Lourtioz, Appl. Phys. Lett. **55**, 82 (1989).
[5] A. M. Samson, S. I. Turovets, V. N. Chizhevskii, and V. V. Churakov, Sov. Phys. JETP **74**, 628 (1992); V. N. Chizhevsky, J. Opt. B: Quantum Semiclass. Opt. **2**, 711 (2000).
[6] T. Mukai and K. Otsuka, Phys. Rev. Lett. **55**, 1711 (1985).
[7] J. Sacher, W. Elässer, and E. Göbel, Phys. Rev. Lett. **63**, 2224 (1989).
[8] J. Ye, H. Li, and J. G. McInerney, Phys. Rev. A **47**, 2249 (1993).
[9] R. Lang and K. Kobayashi, IEEE J. Quantum Electron. **16**, 347 (1980).
[10] M. Sciamanna and K. A. Shore, Nature Photonics **9**, 151 (2015).
[11] G. L. Oppo and A. Politi, Z. Phys. B **59**, 119 (1985).
[12] M. Toda, Phys. Rep. **18**, 1 (1975).
[13] T. Kimura and K. Otsuka, IEEE J. Quantum Electron. **6**, 764 (1970).
[14] S. R. Chinn, H. Y-P. Hong, and J. W. Pierce, IEEE J. Quantum Electron. **QE-12**, 189 (1976).
[15] K. Kubodera and K. Otsuka, IEEE J. Quantum Electron. **17**, 1139 (1981).
[16] S. Tarucha and K. Otsuka, IEEE J. Quantum Electron. **17**, 810 (1981).
[17] J.-L. Chern, K. Otsuka, and F. Ishiyama, Opt. Commun. **96**, 259 (1993); F. Ishiyama, J. Opt. Soc. Am. B **16**, 2202 (1999).
[18] C. Nicolis and G. Nicolis, Tellus **33**, 225 (1981).
[19] G. Giacomelli, M. Giudici, S. Balle, and J. R. Tredicce, Phys. Rev. Lett. **84**, 3298 (2000).
[20] K. Otsuka, Sensors **11**, 2195 (2011).
[21] H. J. Jensen, *Self-organized criticality: emergent complex behavior in physical and biological systems* (Cambridge University Press, 1998).
[22] Per Bak, *How Nature Works: the science of self-organized criticality* (Springer, 2013).
[23] K. Otsuka, IEEE J. Quantum Electron. **15**, 655 (1979).
[24] Donald L. Turcotte, Rep. Prog. Phys. **62**, 1377 (1999).
[25] K. Otsuka, J.-Y. Ko and T. Kubota, Opt. Lett. **26**, 638 (2001).
[26] K Ikeda, K. Otsuka and K. Matsumoto, Prog. Theoretical Phys. (Supplement) **99**, 295 (1989).
[27] K. Otsuka, Phys. Rev. Lett. **65**, 329 (1990).
[28] F. T. Arecchi, G. Giacomelli, P. L. Ramazza, S. Residori, Phys. Rev. Lett. **65**, 2531 (1990).
[29] K. Kaneko and I. Tsuda, Physica D **75**, 1 (1994).
[30] M. Itoh and M. Kimoto, Physica D: Nonlinear Phenomena **109**, 274 (1997).
[31] I. Tsuda, Behavioral and Brain Sciences **24**, 793 (2001).
[32] M. Faggini, B. Bruno and A. Parziale, Journal of the Knowledge Economy **12**, 770 (2021).
[33] L. Cohen, *Time-Frequency Analysis* (Prentice Hall, New York, 1995).
[34] D. G. Carlson, J. Appl. Phys. **39**, 4363 (1968).
[35] A. E. Siegman, *Lasers* (University Science Books, 1986), pp. 1018-1019.
[36] M. Toda, Progress of Theoretical Physics Supplement, **45**, 179 (1970); M. Toda, *Theory of nonlinear lattices* (Springer, 1989).